\documentclass[aoas,MSNbibl,nameyear,seceqn,rotating,dvips]{arximspdf}
\usepackage{dcolumn}
\usepackage[vlined]{algorithm2e}
\usepackage{graphicx,breakurl}

\doi{10.1214/13-AOAS643} 
\volume{7}
\issue{4}
\pubyear{2013}
\firstpage{2431}
\lastpage{2457}

\makeatletter
\newcolumntype{d}[1]{D{.}{.}{#1}}
\let\my@algocf@latexcaption\algocf@latexcaption
\let\my@addcontentsline\addcontentsline
\long\def\algocf@latexcaption#1[#2]#3{%
\def\addcontentsline##1##2##3{}%
\my@algocf@latexcaption{#1}[#2]{#3}%
\global\let\addcontentsline\my@addcontentsline%
}
\makeatother

\begin{document}
\begin{frontmatter}

\title{Multi-way blockmodels for analyzing coordinated
high-dimensional responses\thanksref{T1}}
\runtitle{Multi-way blockmodels}

\begin{aug}
\author[A]{\fnms{Edoardo M.} \snm{Airoldi}\corref{}\ead[label=e1]{airoldi@fas.harvard.edu}},
\author[B]{\fnms{Xiaopei} \snm{Wang}}
\and
\author[C]{\fnms{Xiaodong} \snm{Lin}}
\runauthor{E. M. Airoldi, X. Wang and X. Lin}
\affiliation{Harvard University, University of
Cincinnati and Rutgers University}
\address[A]{E. M. Airoldi\\
Harvard University\\
1 Oxford Street\\
Cambridge, Massachusetts 02138\\
USA\\
\printead{e1}} 
\address[B]{X. Wang\\
University of Cincinnati\\
2815 Commons Way\\
Cincinnati, Ohio 45221\\
USA}
\address[C]{X. Lin\\
Rutgers University\\
100 Rock Avenue\\
Piscataway, New Jersey 08854\\
USA}
\end{aug}
\thankstext{T1}{Supported in part by NSF Grants DMS-11-06980,
IIS-10-17967 and CAREER IIS-11-49662,
and by NIH Grant R01 GM096193 all to Harvard University,
by a faculty research grant from Rutgers Business School, and by a Taft
fellowship to the University of Cincinnati.}

\received{\smonth{12} \syear{2011}}
\revised{\smonth{12} \syear{2012}}

%
\begin{abstract}
We consider the problem of quantifying temporal coordination between
multiple high-dimensional responses. We introduce a family of multi-way
stochastic blockmodels suited for this problem, which avoids
preprocessing steps such as binning and thresholding commonly adopted
for this type of data, in biology. We develop two inference procedures
based on collapsed Gibbs sampling and variational methods. We provide a
thorough evaluation of the proposed methods on simulated data, in terms
of membership and blockmodel estimation, predictions out-of-sample and
run-time. We also quantify the effects of censoring procedures such as
binning and thresholding on the estimation tasks. We use these models
to carry out an empirical analysis of the functional mechanisms driving
the coordination between gene expression and metabolite concentrations
during carbon and nitrogen starvation, in \textit{S.~cerevisiae}.
\end{abstract}

%

\end{frontmatter}

\setcounter{footnote}{1}
\section{Introduction}


In recent years, the biology community at large has engaged in an
effort to characterize \textit{coordinated} mechanisms of cellular
regulation, to enable a \textit{systems-level} understanding of
cellular functions. Reference databases, such as the yeast genome
database (\citeauthor{SGD}), catalog the many regulatory roles of genes and
proteins with links to the originating literature
[\citet{cherballwengjuvi1997}, \citet{kanegoto2000}]. Recent
work spans approaches that leverage these databases to integrate
genomic information across multiple studies and technologies about the
same regulatory mechanism, for example, transcription
[\citet{copezhongarrparm2004}, \citet{Franks2012}], as well
as approaches to integrate genomic information across levels of
regulation, for example, epigenetic markers, chromatin modifications,
transcription and translation [\citet{TroyDoliOwenAltm2003}, \citet{lumarkunwileek2009},
\citet{markairolemitroy2009}].

We consider the problem of quantifying temporal coordination between
gene expression and metabolite concentrations in yeast
[\citeauthor{brauyuanbennlu2006} (\citeyear{brauyuanbennlu2006,BrauHuttAiroRose2007})].
More generally, we are interested in statistical methods to analyze
multiple coordinated high-dimensional measurements about a~system
organism, where correlation among pairs of measurements is believed to
indicate coordinated functional and regulatory roles. We develop
methods for analyzing experiments on regulation dynamics that involve
the following: (1)~data collections about multiple stages of
regulations (transcriptional and metabolic) that offer complementary
views of the cellular response (to Nitrogen and Carbon starvation),
quantified in terms of high-dimensional measurements; and (2)~data
\mbox{collected} according to a specific coordinated temporal design,
whereby the experiments at different stages of regulation are conducted
on cell cultures with matching conditions (nutrient limitations,
environmental stress and chemical compounds present) over time.
Coordinated time courses about complementary stages of regulation
arguably provide the best opportunity to characterize coordinated
regulation dynamics, quantitatively.

A popular approach to study coordinated cellular responses in biology
involves Bayesian networks [\citet{bradbraurabitroy2009},
\citet{TroyDoliOwenAltm2003}]. This approach requires
\textit{binning} real-valued measurements into discrete categories. A
deterministic alternative to explore coordination is the
cross-associations algorithm [\citet{ChakPapaModhFalo2004}], which
instead requires \textit{thresholding} the matrix of correlations
between pairs of genes and metabolites into binary on--off relations.
While binning and thresholding are accepted data preprocessing steps in
the computational biology literature, they raise serious statistical
issues [\citet{blocmeng2103}]. On the one hand, the lack of
appropriate and principled alternatives, together with the sizable
amount of data typical in a coordinated study of cellular responses,
for example, genome-wide expression and hundreds of metabolites, make
preprocessing necessary. These preprocessing steps reduce the
computational burden of the analysis with Bayesian networks and
cross-associations. On the other hand, however, these preprocessing
steps are essentially censoring mechanisms that may compromise the
patterns of variation and covariation in the original data, when the
discovery in such patterns, local and global, is the primary goal of
the analysis [\citet{turn1976}, \citet{vard1985}].

%
%
%

In this paper we develop a family of blockmodels to analyze a
correlation matrix among sets of temporally paired measurements on two
distinct populations of objects. Our work extends a recent block
modeling approach that leverages the notion of \textit{structural
equivalence} [\citet{SnijNowi1997}, \citet{NowiSnij2001}] to
the analysis of coordinated measurements on two populations. For more
details on blockmodels see \citet{goldzhenfienairo2010}.
%
Section~\ref{secmix-pq}\vadjust{\goodbreak} introduces two-way (and multi-way) stochastic
blockmodels for a function of the high-dimensional responses, such as
their correlation. These simple models explicitly allow different
objects in the two (or more) populations to be associated with multiple
blocks, say, of correlation, to different degrees, and does not require
binning or thresholding.
Estimation and inference using variational methods is outlined in
Section~\ref{secest-inf}. Details of variational and MCMC inference are
provided in the supplement [\citet{Airoldi2013fk}].
Section~\ref{secemp-simu} develops a thorough evaluation of the
proposed methods on simulated data, including a comparative evaluation
of the MCMC and variational inference procedures in terms of the
following: (1)~membership and blockmodel matrix estimation,
(2)~predictions out-of-sample, and (3)~run-time. We assess the effects
of thresholding on inference in Section~\ref{secsimcensoring}. In
Section~\ref{secmet} we analyze two recently published collections of
time-course data to explore the functional mechanisms underlying the
coordination of transcription and metabolism during carbon and nitrogen
starvation, in \textit{S. cerevisiae}. We compare the results with
published results on the same data using binning and Bayesian networks,
and to new results we obtain using thresholding and
cross-associations.\vspace*{-3pt}

\section{Multi-way stochastic blockmodels}
\label{secmix-pq}

In this section we introduce multi-way stochastic blockmodels and the
associated inference procedures. This family of models generalizes
mixed membership stochastic blockmodels for analyzing interactions
within a single population [\citet{AiroBleiFienXing2008}] to
interactions between two or more populations. Multi-way stochastic
blockmodels models enable the discovery of interactions between latent
groups across different populations, and provide estimates of the group
memberships for each subject. We develop two inference strategies: one
based on collapsed Gibbs sampling [\citet{liu1994}], the other
based on variational Expectation--Maximization (vEM)
[\citet{JordGhahJaakSaul1999}, \citet{Airo2007b}].\vspace*{-3pt} 


\subsection{Two-way blockmodels}\label{sec2model}
Consider a two-way interaction table between two sets of nodes
$\mathcal{N}_1$ and $\mathcal{N}_2$ of size $N_1$ and $N_2$,
respectively. These two sets of nodes represent elements of two
distinct populations. An observation $Y(j,k)$, $j=1,\ldots,N_1$,
$k=1,\ldots,N_2$, denotes the strength of the interaction between the
$j$th element of $\mathcal{N}_1$ and the $k$th element of
$\mathcal{N}_2$.

As a running example, we consider the coordinated time course data we
analyze in Section~\ref{secmet}. The data consists of $N_1$ time series
of gene expression levels and of $N_2$ time series of metabolite
concentrations, before and after Nitrogen and Carbon starvation for a
total of seven time points, in yeast [\citet{brauyuanbennlu2006},
\citet{bradbraurabitroy2009}]. We posit a model for the $N_1
\times N_2$ matrix of Fisher-transformed correlations of time courses
for each gene--metabolite pair or for any of its sub-matrices obtained
by selecting subsets of genes and metabolites of special interest to
biologists. The goal of the analysis is to reveal interactions between
gene functions and metabolic pathways, operationally defined as sets of
genes and sets of metabolites, respectively, with similar correlation
patterns.\vadjust{\goodbreak}

In the context of this application, we posit that each gene can
participate in up to $K_1$ functions, that is, latent row groups, and
that each metabolite can participate in up to $K_2$ metabolic pathways,
that is, latent column groups.\footnote{We refer to gene functions and
metabolic pathways as defined in the yeast genome database and the
Kyoto encyclopedia of genes and genomes.} Latent Dirichlet vectors
$\vec\pi_j$ and $\vec p_k$ capture the relative fractions of time gene
$j$ and metabolite $k$~participate in the different cellular functions
and pathways, or latent groups. The distribution of the correlation,
or, more generally, interaction, $Y(j,k)$, is then a~function of the
interactions among the latent groups, fully specified by a $K_1 \times
K_2$ matrix $B$, together with the latent memberships of the gene and
metabolite involved.
The data generating process, given $\alpha,\beta,B$ and $\sigma$, is
as follows:
%
\begin{eqnarray}
\vec{\pi}_j & \sim& \operatorname{Dirichlet} (\alpha),
\\
\vec{p}_k & \sim&\operatorname{Dirichlet} (\beta),
\\
Y(j,k) & \sim&\operatorname{Normal} \bigl(\vec{\pi}_j'
B \vec{p}_k,\sigma^2 \bigr),
\end{eqnarray}
%
%
where indices $j=1,\ldots,N_1$ and $k=1,\ldots,N_2$ run over genes and
metabolites, respectively, vectors $\vec{\pi}_j$ and $\vec{p}_k$ are
$K_1$- and $K_2$-dimensional, respectively, and elements of the
blockmodel mean matrix $B_{gh} \in\mathbb{R}$.

While the observations $Y(j,k)$ in the motivation application are
Fisher-transformed correlations, real-valued with real-valued mean
matrix $B$, the proposed models are more flexible. For instance, we
develop a two-way block model for binary observations in
Section~\ref{secngmodel}, that is used in Section~\ref{secsimcensoring}
for quantifying the effects of censoring the data matrix $Y$.

For inference purposes, we consider an augmented data generating
process, in which we introduce latent indicator vectors $\vec
{D}_{j\rightarrow k}$ and $\vec{E}_{j \leftarrow k}$ that denote the
single memberships of gene $j$ and metabolite $k$ for the correlation
$Y(j,k)$. The latent indicators $\{D, E\}$ do not have a clear
biological interpretation, but serve to improve computational
tractability of the inference; they lead to optimization problems that
have analytical solutions. The trade-offs of such a strategy have been
explored elsewhere [e.g., see \citet{AiroBleiFienXing2008}]. From
a statistical perspective, introducing $\{D, E\}$ amounts to a specific
representation of the interactions in terms of random effects.

\subsection{Extension to non-Gaussian responses}\label{secngmodel}
In the data generating process above, $Y$ is generated from a Normal
distribution and the blockmodel's elements take real values. Extending
the proposed model to other distributions to account for data $Y$ that
live in a different space is straightforward. And because of the
hierarchical structure of the model, only a minor portion of the
inference and estimation strategies detailed in
Section~\ref{secest-inf} will need to be modified appropriately, as a
consequence.\vadjust{\goodbreak}



We will consider one such extension to binary observations
$Y(j,k)$---namely, correlations after thresholding---in
Section~\ref{secsimcensoring} to assess the effects of preprocessing on
the accuracy in estimating the blockmodel.
The data generating process in Section~\ref{sec2model} is modified as
follows.
The blockmodel's elements now take values in the unit interval, since
they capture the probability that there is a correlation above
threshold between members of any pair of blocks, $B_{gh}\in[0,1]$. For
each pair $(j,k)$, $j=1,\ldots,N_1$, $k=1,\ldots,N_2$, we\vspace*{1pt}
sample the pairwise binary observation $Y(j,k)
\sim\operatorname{Bernoulli} (\vec{D}_{j\rightarrow k}' B \vec{E}_{j
\leftarrow k})$. Variational Bayes and MCMC inference also remain
mostly unchanged. New updating equations for the elements of $B$ will
be needed; see equation (\ref{VB5}) and the supplement
[\citet{Airoldi2013fk}].

\subsection{Extension to multi-way blockmodels}\label{secmmodel}
The two-way blockmodel introduced above can also be extended for
analyzing multi-way interactions between three or more
populations.

Consider a three-way interaction table $Y(i_1,i_2,i_3)$ observed on
three populations $\mathcal{N}_1$, $\mathcal{N}_2$, $\mathcal{N}_3$,
where $i_1 \in\mathcal{N}_1$, $i_2 \in \mathcal{N}_2$ and $i_3
\in\mathcal{N}_3$. Assume that there are $K_1$, $K_2$ and $K_3$ latent
groups existing in $\mathcal{N}_1$, $\mathcal{N}_2$ and $
\mathcal{N}_3$, respectively. We can treat the three way interaction
observed in $Y$ as a result of three way group interactions. Namely,
$Y(i_1,i_2,i_3)$ can be fully characterized by $B(g_1,g_2,g_3)$, with
items $\{i_1,i_2,i_3\}$ belonging to group $\{g_1,g_2,g_3\}$,
respectively. Therefore, inferences procedures for this three-way
blockmodel can be developed in a similar fashion as those for the
two-way blockmodel. Note that although the ideas for generalizations to
higher order tables remain the same, keeping track of indices during
inference becomes tedious.


%

\subsection{Parameter estimation and posterior inference}\label{secest-inf}
The main inference task is to estimate the matrix $B$ and the mixed
membership vectors $\vec{\pi}$ and $\vec{p}$. Given the observed data
$Y=Y(j,k)$, latent variable
$X=\{\vec{\pi}_j,\vec{p_k},\vec{D}_{j\rightarrow k},\vec{E}_{j
\leftarrow k}\}$ and the parameters $\Theta=\{\alpha, \beta, \sigma^2,
B\}$, the complete data likelihood $p(Y,X|\Theta)$ can be written~as
%
\begin{eqnarray}
\qquad&& \label{2} p \bigl(Y,X|{\alpha},{\beta},B,\sigma^2 \bigr)
\nonumber
\\
&&\qquad = \prod_{j}p_1(\vec{\pi
}_j| {\alpha})\prod_{k}p_1(
\vec{p}_k|{\beta})
\\
&&\quad\qquad{}\times \prod_{j,k}p_0 \bigl(Y(j,k)|
\vec{D}_{j\rightarrow
k},\vec{E}_{j \leftarrow k},B,\sigma^2
\bigr)p_2(\vec{D}_{j\rightarrow
k}|\vec{\pi}_j)p_2(
\vec{E}_{j \leftarrow k}|\vec{p_k}),\nonumber
\end{eqnarray}
where $p_0$ is a Normal distribution with mean
$\mu=\vec{D}_{j\rightarrow k}' B \vec{E}_{j \leftarrow k}$ and
variance $\sigma^2$, $p_1$ is a Dirichlet distribution, and $p_2$
is a Multinomial distribution with $n=1$. The
posterior distribution of the latent variable $X$ is
%
\begin{equation}
p(X|Y,\Theta)=\frac{p(Y,X|\Theta)}{p(Y|\Theta)},\vadjust{\goodbreak}
\end{equation}
where
the marginal distribution $p(Y|\Theta)$ has the following form:
\begin{eqnarray*}
p(Y|\Theta)&=&\int_{X}p(Y,X|\Theta)\,dX
\\[-2pt]
&=&\sum_{\vec{D}}\sum_{\vec{E}}
\biggl\{\int\!\!\int \prod_{j}p_1(\vec{
\pi}_j|\alpha)\prod_{k}p_1(\vec{p}_k|\beta )
\\[-2pt]
&&\hspace*{52pt}{}\times \prod_{j,k}p_2(\vec{D}_{j\rightarrow k}|\vec{\pi}_j)p_2(\vec{E}_{j \leftarrow k}|\vec{p_k})\,d\vec{\pi}\,d\vec{p}
\\[-2pt]
&&\hspace*{61pt}{}\times \prod_{j,k}p_0 \bigl(Y(j,k)|\vec
{D}_{j\rightarrow k},\vec{E}_{j \leftarrow k},B,\sigma^2 \bigr) \biggr\}.
\end{eqnarray*}

There does not exist an explicit solution to the maximization of
$p(Y | \Theta)$. Therefore, we propose an iterative procedure based
on variational Bayes for parameter estimation. In comparison, we
also develop a MCMC scheme based on collapsed Gibbs sampling to
achieve the desired statistical inferences.

\subsubsection{Variational expectation--maximization}

To achieve variational inference, we introduce free variational
parameters $\vec{\nu}_{j}$ and $\vec{\xi}_{k}$ to approximate $\vec
{\pi}_j$ and $\vec{p_k}$, free variational variables $\vec{\phi
}_{j\rightarrow k}$ and $\vec{\eta}_{j\leftarrow k}$ to approximate
$\vec{D}_{j\rightarrow k}$ and $\vec{E}_{j \leftarrow k}$, and latent
distribution $q(X)$ to approximate the true posterior distribution
$p(X|Y,\Theta)$. By Jensen's inequality, we have the following
likelihood lower bound:
%
\begin{equation}\label{jesen1}
\log p(Y|\Theta)\geq E_q \bigl[\log p(Y,X|\Theta)
\bigr]-E_q \bigl[\log q(X) \bigr].
\end{equation}

A coordinate ascend algorithm can be applied to obtain a local
maximizer of this lower bound, which results in the updates
(\ref{VB1})--(\ref{VB5}). Detailed derivations are left in the
supplementary material [\citet{Airoldi2013fk}]. The resulting variational EM algorithm is given in
Algorithm~\ref{algoVB}:
%
\begin{eqnarray}
\phi_{j\rightarrow k,g} &\propto& \exp \biggl({\psi(\nu_{j,g})-\psi \biggl(
\sum_g\nu_{j,g} \biggr) \biggr)}
\nonumber\\[-9pt]\label{VB1} \\[-9pt]
&&{}\times \prod _h \bigl(\sigma ^2\cdot e^{(Y(j,k)-B(g,h) )^2/\sigma^2}
\bigr)^{-1/2\eta_{j\leftarrow k,h}},\nonumber
\\[-2pt]
\eta_{j\leftarrow k,h} &\propto& \exp{ \biggl(\psi(\xi_{k,h})-\psi \biggl(
\sum_h\xi_{k,h} \biggr) \biggr)}
\nonumber\\[-9pt]\label{VB2} \\[-9pt]
&&{}\times \prod_g \bigl(\sigma ^2\cdot e^{(Y(j,k)-B(g,h))^2/\sigma^2}
\bigr)^{-1/2\phi_{j\rightarrow k,g}},\nonumber
\\[-2pt]
\nu_{j,g}&=& \sum_k\phi_{j\rightarrow k,g}+\alpha, \label{VB3}
\\[-2pt]
\xi_{k,h}&=& \sum_j\eta_{j\leftarrow k,h}+\beta, \label{VB4}
\\[-2pt]
B(g,h)&=&\frac{\sum_{j,k}\phi_{j\rightarrow k,g}\eta_{j\leftarrow
k,h}Y(j,k)}{\sum_{j,k}\phi_{j\rightarrow k,g}\eta_{j\leftarrow k,h}}.\label{VB5}
\end{eqnarray}

\section{Evaluating inference and effects of preprocessing}\label{secemp-simu}
Here we use simulated data to compare the performance of
variational and MCMC inference procedures for the two-way block
model along multiple dimensions: estimation accuracy of mixed
membership vectors, accuracy of predictions out-of-sample,
\mbox{estimation} accuracy of the blockmodel interaction matrix $B$ and
run-time. This extensive comparative evaluation provides a
practical guideline for choosing the proper inference procedure in
a real setting, especially when analyzing large tables. In
addition, we quantify the effect of censoring on the inference in
terms of estimation error.

\begin{algorithm}[t]
\caption{The variational EM algorithm. The E steps 6--9 are also
repeated until convergence to achieve the most stabilized mutual
updates for the set of\vspace*{1pt} free parameters $\vec{\phi}$, $\vec{\eta}$,
$\vec{\nu}$, $\vec{\xi}$.
}\label{algoVB}
\begin{minipage}{342pt}
{\fontsize{9.8pt}{13pt}\selectfont{\textbf{Variational EM}
$\bigm(Y(j,k)_{j=1,k=1}^{N_1,N_2}, \alpha,
\beta, \sigma^2 \bigm)$\\
\smallskip
\hspace*{2pt}\nl initialize $\vec{\phi}_{j\rightarrow k}:= 1/K_1$ for all $j$ and $k$ \\[-6pt]
\hspace*{2pt}\nl initialize $\vec{\eta}_{j\leftarrow k}:= 1/K_2$ for all $j$ and $k$ \\
\hspace*{2pt}\nl initialize $\vec{\nu}_{j}:= N_2/K_1+\alpha$ for all $j$ \\
\hspace*{2pt}\nl initialize $\vec{\xi}_{k}:= N_1/K_2+\beta$ for all $k$ \\
\hspace*{2pt}\nl initialize $B(g,h)$ for all $g$ and $h$ as the data mean plus a random noise \\
\Repeat{\mbox{\rm convergence}}{

\hspace*{2pt}\nl E step: update $\vec{\phi}_{j\rightarrow k}$ for all
$j$ and $k$
using equation (\ref{VB1}) and normalize to sum to $1$ \\
\hspace*{2pt}\nl\hspace{1.1cm}update $\vec{\eta}_{j\leftarrow k}$ for
all $j$
and $k$ using equation (\ref{VB2}) and normalize to sum to $1$ \\
\hspace*{2pt}\nl\hspace{1.1cm}update $\vec{\nu}_{j}$ for all $j$ using
equation
(\ref{VB3}) \\
\hspace*{2pt}\nl\hspace{1.1cm}update $\vec{\xi}_{k}$ for all $k$ using
equation
(\ref{VB4}) \\
\hspace*{2pt}\nl M step: update $B(g,h)$ for all $g,h$ using equation (\ref{VB5})\\[-10pt]
} \hspace*{2pt}\nl\hspace{-2pt}\textbf{return}
$(\vec{\phi},\vec{\eta},\vec{\nu},\vec{\xi},B)$}}
\smallskip
\end{minipage}
\end{algorithm}

\subsection{Design of experiments}\label{secsimu-setup}
In the past decade, variational EM (vEM) has become a practical
alternative to MCMC when dealing with large data sets, despite its lack
of theoretical guarantees [\citet{JordGhahJaakSaul1999},
\citet{Airo2007b}, \citet{JoutAiroFienLove2007}]. The
relative merits between vEM and MCMC have been established empirically
for a number of models [e.g., see \citet{BleiJord2006},
\citet{braumcau2010}]. We designed simulations with the goal of
exploring the trade-off between estimation accuracy and computational
burden that vEM helps manage in the context of estimation and posterior
inference with the proposed model.

Briefly, vEM is an optimization approach, no sampling is involved,
which requires key choices about the following: (1) error tolerance for
both the approximate E step and the M step, and (2) how to design
multiple initializations and how many to use. MCMC is a sampling
approach, which requires key choices about the following: (1)
convergence criteria, (2) burn-in, (3) thinning to reduce
autocorrelation, and (4) multiple chains.
For the variational EM approach, we set the overall error tolerance at
1e--5, the maximum number of iterations for the variational E steps at
10, and 10 random initializations.
For the MCMC approach, we investigated the convergence using
Gelman--Rubin and Raftery--Lewis for the median, autocorrelation using
trace plots and partial autocorrelation functions. Based on these
studies, we chose to use 1000 iterations for burn-in, 6000 iterations
and a 10~to~1 thinning ratio, which results in 500 draws for each
chain, and we used 10 chains. 
%
For both approaches, we use the true Dirichlet parameters $\alpha,
\beta$ and the true variance $\sigma^2=0.01$. Overall, this seems a
fair comparison.

The data are generated using the procedures described in
Section~\ref{sec2model} with the following specifications. The $B(g,h)$
follows a Normal distribution $B(g,h) \sim \operatorname{Normal}(\mu_B(g,h),\sigma^2_B(g,h))$, where $\mu_B(g,h)=0$ and
$\sigma^2_B(g,h)=1$. Three sets of block sizes are considered:
$(K_1,K_2)=(2,3)$, $(4,6)$ and $(6,9)$. The corresponding table sizes
are $(N_1, N_2)= (10,15)$, $(50, 75)$ and $(100,150)$, respectively.
The Dirichlet parameters are set to be $\alpha=\beta=0.2$ or
$\alpha=\beta=0.05$. In all the experiments, we set $\sigma^2=0.01$.



\subsection{Mixed membership estimation}\label{secacc-eff}
Here we evaluate the competing estimation procedures on recovering
mixed membership vectors.
We report results on the accuracy of the first and second largest
membership components.
It is well known that mixture models and mixed membership models suffer
from identifiability issues, that is, their likelihood is uniquely
specified up to permutations of the labels
[\citet{tittsmitmako1985}]. We evaluate the performance for a
fixed permutation, obtained empirically by sorting the membership
vectors for the vEM and by using a standard Procrustes transform for
the MCMC [\citet{Step2000}].
We note that vEM converged quickly to a (local) optimum, thus involving
a considerably more mitigated label switching issue than the collapsed
Gibbs sampler. This is an advantage, especially given that the
empirical vEM estimation error reported in
Table~\ref{tabmemVB-dBdecreavg} is comparable to that of the more
principled MCMC sampler.

\begin{sidewaystable}
{\fontsize{9pt}{10.7pt}\selectfont{
\tabcolsep=0pt
\textwidth=\textheight
\tablewidth=\textwidth
\caption{Comparisons on row and column estimation
accuracy of estimates for the first highest membership (regular font)
and second highest membership (italic font) obtained with variational
EM~and~MCMC. Standard errors are quoted inside parenthesis}\label{tabmemVB-dBdecreavg}
\begin{tabular*}{\tablewidth}{@{\extracolsep{\fill}}ld{1.2}c@{\hspace*{-25pt}}cc@{\hspace*{-25pt}}cc@{\hspace*{-25pt}}c@{}}
\hline
& & \multicolumn{2}{c}{$\bolds{K_1=2}$ \textbf{and} $\bolds{K_2=3}$}
  & \multicolumn{2}{c}{$\bolds{K_1=4}$ \textbf{and}  $\bolds{K_2=6}$}
  & \multicolumn{2}{c@{}}{$\bolds{K_1=6}$ \textbf{and}  $\bolds{K_2=9}$}\\[-6pt]
& & \multicolumn{2}{c}{\hrulefill}
  & \multicolumn{2}{c}{\hrulefill}
  & \multicolumn{2}{c@{}}{\hrulefill}
 \\
\textbf{($\bolds{N_1}$, $\bolds{N_2}$)} & \multicolumn{1}{c}{$\bolds{\alpha/\beta}$} &
\textbf{Row} & \textbf{Column} & \textbf{Row} & \textbf{Column} & \textbf{Row} & \textbf{Column} \\
\hline
vEM & &&&\\
$(10, 15)$ & 0.2 & 0.970 (0.067) & 0.667 (0.031) & 0.620 (0.063) & 0.587 (0.069) &
0.470 (0.048) & 0.520 (0.076) \\ 
& & \textit{0.970} (\textit{0.067}) & \textit{0.522} (\textit{0.075}) & \textit{0.233} (\textit{0.152}) & \textit{0.179} (\textit{0.077}) &
\textit{0.210} (\textit{0.129}) & \textit{0.060} (\textit{0.058}) \\ 
& 0.05 & 0.980 (0.042) & 0.967 (0.085) & 0.870 (0.125) & 0.807 (0.066) &
0.780 (0.063) & 0.567 (0.085) \\ 
& & \textit{0.980} (\textit{0.042}) & \textit{0.533} (\textit{0.233}) & \textit{0.233} (\textit{0.179}) & \textit{0.190} (\textit{0.110}) &
\textit{0.317} (\textit{0.123}) & \textit{0.133} (\textit{0.112})
\\[3pt] 
$(50, 75)$ & 0.2 & 0.784 (0.122) & 0.751 (0.146) & 0.680 (0.034) & 0.471 (0.039) &
0.426 (0.053) & 0.416 (0.041) \\ 
& & \textit{0.784} (\textit{0.122}) & \textit{0.694} (\textit{0.130}) & \textit{0.304} (\textit{0.097}) & \textit{0.175} (\textit{0.033}) &
\textit{0.194} (\textit{0.054}) & \textit{0.136} (\textit{0.031}) \\ 
& 0.05 & 0.980 (0.000) & 0.849 (0.074) & 0.620 (0.104) & 0.575 (0.058) &
0.634 (0.046) & 0.483 (0.053) \\ 
& & \textit{0.980} (\textit{0.000}) & \textit{0.662} (\textit{0.132}) & \textit{0.239} (\textit{0.118}) & \textit{0.216} (\textit{0.058}) &
\textit{0.210} (\textit{0.073}) & \textit{0.149} (\textit{0.047})
\\[3pt] 
$(100, 150)$ & 0.2 & 0.960 (0.000) & 0.823 (0.106) & 0.601 (0.077) & 0.670 (0.076)
& 0.485 (0.048) & 0.357 (0.029) \\ 
& & \textit{0.960} (\textit{0.000}) & \textit{0.612} (\textit{0.247}) & \textit{0.261} (\textit{0.055}) & \textit{0.237} (\textit{0.063}) &
\textit{0.194} (\textit{0.029}) & \textit{0.137} (\textit{0.022}) \\ 
& 0.05 & 0.946 (0.092) & 0.743 (0.132) & 0.769 (0.055) & 0.707 (0.057) &
0.553 (0.084) & 0.479 (0.052) \\ 
& & \textit{0.946} (\textit{0.092}) & \textit{0.520} (\textit{0.227}) &
\textit{0.361} (\textit{0.057}) & \textit{0.236} (\textit{0.064}) &
\textit{0.217} (\textit{0.060}) & \textit{0.135} (\textit{0.028})\\[3pt] 
MCMC & &&&\\
$(10, 15)$ & 0.2 & 0.922 (0.148) & 0.730 (0.102) & 0.678 (0.015) & 0.665 (0.012) &
0.669 (0.008) & 0.521 (0.007) \\ 
& & \textit{0.922} (\textit{0.148}) & \textit{0.504} (\textit{0.167}) & \textit{0.306} (\textit{0.053}) & \textit{0.204} (\textit{0.031}) &
\textit{0.207} (\textit{0.011}) & \textit{0.157} (\textit{0.004}) \\ 
& 0.05 & 0.841 (0.121) & 0.901 (0.120) & 1.000 (0.000) & 0.878 (0.031) &
0.884 (0.005) & 0.825 (0.007) \\ 
& & \textit{0.841} (\textit{0.121}) & \textit{0.409} (\textit{0.138}) & \textit{0.520} (\textit{0.122}) & \textit{0.413} (\textit{0.091}) &
\textit{0.227} (\textit{0.052}) & \textit{0.161} (\textit{0.022})
\\[3pt] 
$(50, 75)$ & 0.2 & 0.871 (0.121) & 0.671 (0.097) & 0.711 (0.095) & 0.659 (0.084) &
0.682 (0.106) & 0.562 (0.051) \\ 
& & \textit{0.871} (\textit{0.121}) & \textit{0.437} (\textit{0.186}) & \textit{0.380} (\textit{0.039}) & \textit{0.300} (\textit{0.065}) &
\textit{0.301} (\textit{0.093}) & \textit{0.231} (\textit{0.026}) \\ 
& 0.05 & 0.994 (0.013) & 0.676 (0.113) & 0.775 (0.176) & 0.753 (0.129) &
0.824 (0.088) & 0.839 (0.054) \\ 
& & \textit{0.994} (\textit{0.013}) & \textit{0.452} (\textit{0.131}) & \textit{0.383} (\textit{0.135}) & \textit{0.319} (\textit{0.142}) &
\textit{0.357} (\textit{0.090}) & \textit{0.365} (\textit{0.074})
\\[3pt] 
$(100, 150)$ & 0.2 & 0.971 (0.032) & 0.653 (0.150) & 0.682 (0.119) & 0.633 (0.083)
& 0.735 (0.069) & 0.614 (0.078) \\ 
& & \textit{0.968} (\textit{0.034}) & \textit{0.420} (\textit{0.223}) & \textit{0.332} (\textit{0.080}) & \textit{0.255} (\textit{0.054}) &
\textit{0.310} (\textit{0.074}) & \textit{0.235} (\textit{0.059}) \\ 
& 0.05 & 0.830 (0.208) & 0.773 (0.138) & 0.810 (0.140) & 0.772 (0.127) &
0.780 (0.046) & 0.750 (0.064) \\ 
& & \textit{0.829} (\textit{0.208}) & \textit{0.463} (\textit{0.203}) &
\textit{0.354} (\textit{0.151}) & \textit{0.277} (\textit{0.088}) &
\textit{0.285} (\textit{0.053}) & \textit{0.249} (\textit{0.046})\\ 
\hline
\end{tabular*}}}\vspace*{-8pt}
\end{sidewaystable}

To quantify accuracy, we identify the locations of the largest two
components in the estimated vector of probabilities, $\vec {\pi}_j$,
and take those to be the first and second choice of group memberships
for the $j$th row. These assignments are compared, via zero-one loss,
with the true memberships: if there is a match, we note the accuracy as
1, otherwise 0. The recorded row accuracy is the average over all the
rows and the ten experiments. The column accuracy is defined in a
similar fashion.

The results for the estimated first and second memberships are
summarized in Table~\ref{tabmemVB-dBdecreavg}. The results for the
first membership suggest that estimation is well behaved in the
proposed model; the true membership can be recovered with a fairly high
successful rate under different experimental settings. As expected, the
estimation accuracy decreases with the increase on the block size. The
lowest pair reported in the table are 0.485 and 0.357 for $K_1=6$ and
$K_2=9$, still much better than random assignments where the accuracy
would be $1/6$ and $1/9$, respectively.
For the second membership, we only consider elements with an estimated
second membership probability greater than a threshold. In this study,
the thresholds are $\frac{1}{10K_1}$ and $\frac{1}{10K_2}$ for row and
column memberships, respectively. It is clear that the variational
Bayes approach performs much better than MCMC in estimating the second
membership. One explanation can be that the second membership is more
ambiguous than the first membership, requiring a large number of
iterations for MCMC to converge.

Another factor that affects model performances is the Dirichlet
parameters $\alpha$ and $\beta$. Judging from the table, the
accuracy when $\alpha=\beta=0.05$ is generally higher than those
of $\alpha=\beta=0.2$. This result is reasonable since a smaller
$\alpha$ and $\beta$ value corresponds to a higher likelihood of a
dominating component, which is easier to identify than more
ambiguous memberships.

The membership accuracy computed through variational Bayes aligns
with those calculated from MCMC, and even slightly better when the
block size is small. Since variational inference is typically much
more efficient than MCMC, the former method is preferred for
practical analysis, especially for high-dimensional cases. We will
present run-time comparisons between these two approaches in the
next section.

\begin{sidewaystable}
\tabcolsep=0pt
\textwidth=\textheight
\tablewidth=\textwidth
\caption{Comparisons on row and column estimation
accuracy between variational EM and MCMC, when $2/9$ of the entries are
missing. Standard errors are inside the parenthesis}\label{tabmem3}
\begin{tabular*}{\tablewidth}{@{\extracolsep{\fill}}ld{1.2}c@{\hspace*{-25pt}}cc@{\hspace*{-25pt}}cc@{\hspace*{-25pt}}c@{}}
\hline
& & \multicolumn{2}{c}{$\bolds{K_1=2}$ \textbf{and} $\bolds{K_2=3}$}
  & \multicolumn{2}{c}{$\bolds{K_1=4}$ \textbf{and}  $\bolds{K_2=6}$}
  & \multicolumn{2}{c@{}}{$\bolds{K_1=6}$ \textbf{and}  $\bolds{K_2=9}$}\\[-6pt]
& & \multicolumn{2}{c}{\hrulefill}
  & \multicolumn{2}{c}{\hrulefill}
  & \multicolumn{2}{c@{}}{\hrulefill}\\

\textbf{($\bolds{N_1}$, $\bolds{N_2}$)} & \multicolumn{1}{c}{$\bolds{\alpha/\beta}$} &
\textbf{Row} & \textbf{Column} & \textbf{Row} & \textbf{Column} & \textbf{Row} & \textbf{Column} \\
\hline
vEM & &&&\\
$(10, 15)$ & 0.2 & 0.780 (0.148) & 0.600 (0.094) & 0.610 (0.110) & 0.507 (0.118) &
0.520 (0.063) & 0.547 (0.103) \\
& 0.05 & 0.900 (0.067) & 0.853 (0.117) & 0.730 (0.125) & 0.547 (0.108) &
0.700 (0.094) & 0.613 (0.042)
\\[3pt]
$(50, 75)$ & 0.2 & 0.664 (0.067) & 0.615 (0.134) & 0.452 (0.081) & 0.383 (0.039) &
0.366 (0.034) & 0.335 (0.037)\\
& 0.05 & 0.930 (0.034) & 0.843 (0.077) & 0.570 (0.135) & 0.564 (0.074) &
0.504 (0.076) & 0.444 (0.040)
\\[3pt]
$(100, 150)$ & 0.2 & 0.786 (0.091) & 0.672 (0.128) & 0.472 (0.124) & 0.362 (0.055)
& 0.313 (0.049) & 0.326 (0.059) \\
& 0.05 & 0.751 (0.194) & 0.749 (0.136) & 0.656 (0.102) & 0.503 (0.091) &
0.397 (0.068) & 0.373 (0.056)
\\[3pt]
MCMC &&&&\\
$(10, 15)$ & 0.2 & 0.703 (0.100) & 0.617 (0.083) & 0.480 (0.091) & 0.460 (0.085) &
0.406 (0.012) & 0.368 (0.054)\\
& 0.05 & 0.770 (0.145) & 0.726 (0.115) & 0.540 (0.161) & 0.446 (0.073) &
0.454 (0.101) & 0.456 (0.070)
\\[3pt]
$(50, 75)$ & 0.2 & 0.788 (0.145) & 0.645 (0.104) & 0.544 (0.064) & 0.443 (0.032) &
0.357 (0.062) & 0.343 (0.039) \\
& 0.05 & 0.809 (0.194) & 0.647 (0.098) & 0.606 (0.072) & 0.567 (0.068) &
0.473 (0.054) & 0.479 (0.074)
\\[3pt]
$(100, 150)$ & 0.2 & 0.813 (0.103) & 0.576 (0.102) & 0.575 (0.061) & 0.492 (0.042)
& 0.411 (0.048) & 0.395 (0.028) \\
& 0.05 & 0.867 (0.150) & 0.834 (0.111) & 0.639 (0.051) & 0.524 (0.040) &
0.514 (0.092) & 0.497 (0.030) \\
\hline
\end{tabular*}
\end{sidewaystable}

\subsection{Predictions out-of-sample}\label{secsimoos}
Prediction power is a useful criterion for evaluating statistical
models. When some data are missing, is the model sufficiently flexible
to provide correct inferences and to predict the missing values with
high accuracy? To answer this question, we randomly select $2/3$ of rows
and $2/3$ of columns from the table, whose intersections are $4/9$ of the
entries. We set half of them (i.e., $2/9$) as missing (to avoid
eliminating an entire row or column), and run the model on the
remaining $7/9$ entries. The first membership prediction accuracy is
reported in Table~\ref{tabmem3}. They are slightly lower than those
estimated without missing values, but overall much better than the
baseline probabilities $1/K_1$ and $1/K_2$. Furthermore, the prediction
accuracy achieved by variational Bayes is comparable or better than
those obtained by MCMC. This result reinforces our belief that
variational Bayes is a good inference approach for the proposed
blockmodel.

\subsection{Blockmodel matrix estimation}
Here we compare the variational\break  Bayes and MCMC in terms of estimating
the matrix $B$. The estimation error $\varepsilon_B$ is defined as the
1-norm of the matrix $|B-\hat{B}|$, where $\hat{B}$ is the estimated
matrix. The result for $K_1=2$, $K_2=3$ is shown in Table~\ref{tabB1}.
Except for the case of $\alpha=\beta=0.05$ and $N_1=10$, $N_2=15$,
variational Bayes performs close to or better than MCMC. The true $B$
in this simulation study is
\[
\pmatrix{ -0.5009 & 0.0687 & 1.5887
\cr
0.4148 & -0.8086 & -1.3112}.
\]
%

\begin{table}
\tabcolsep=2pt
\caption{Comparisons on $\varepsilon_B$ as the estimation
error of $B$ between variational Bayes and MCMC} \label{tabB1}
{\fontsize{8.5pt}{\baselineskip}\selectfont
\begin{tabular*}{\textwidth}{@{\extracolsep{\fill}}@{}lccccccc@{}}
\hline
\textbf{($\bolds{N_1}$, $\bolds{N_2}$)} & \multicolumn{2}{c}{\textbf{(10,15)}}
& \multicolumn{2}{c}{\textbf{(50,75)}}  & \multicolumn{2}{c}{\textbf{(100,150)}} \\[-6pt]
& \multicolumn{2}{c}{\hrulefill} & \multicolumn{2}{c}{\hrulefill}  & \multicolumn{2}{c}{\hrulefill} \\
$\bolds{\alpha/\beta}$ & \textbf{0.05} & \textbf{0.2} & \textbf{0.05} & \textbf{0.2} & \textbf{0.05} & \textbf{0.2} \\
\hline
VB &0.152 (0.042) & 0.022 (0.022) &0.048 (0.024) &0.061 (0.061) &0.053 (0.029) & 0.002 (0.001)\\
MCMC &0.019 (0.006) &0.027 (0.047) &0.110 (0.058) &0.045 (0.066) &0.134 (0.065) &0.105 (0.058)\\
\hline
\end{tabular*}}
\end{table}

\subsection{Sensitivity to initialization and priors specifications}\label{secsiminitprior}
Here we analyzed the sensitivity of the inference to informative versus
noninformative prior specifications, and to uniform versus random
initialization of some constants in our model. The results show no
significant sensitivity of the estimation error to these choices. This
evidence supports our claim that inference is well behaved and that
identifiability is not an issue for the model we proposed, in practice,
in the data regimes we considered.

In Algorithm \ref{algoVB} (vEM) and the supplement (MCMC), we
initialized a subset of parameters ($\pi,\eta,\nu,\varepsilon$ in vEM
and $D,E$ in MCMC) uniformly. To assess the sensitivity of inference to
this initialization strategy, we tested alternative versions of these
algorithms in which we initialized these parameters at random, on the
data set analyzed in Section~\ref{secsimcensoring}. Briefly, in vEM, we
initialized each $\vec{\phi}_{j\rightarrow k}$ and
$\vec{\eta}_{j\leftarrow k}$ with random membership vectors, then
initialized $\vec{\nu}_j$, $\vec{\xi}_k$ using equations
(\ref{VB3})~and~(\ref{VB4}). The blockmodel $B$ is initialized as in
Algorithm \ref{algoVB}. In MCMC, we initialized each
$\vec{D}_{j\rightarrow k}$, $\vec{E}_{j \leftarrow k}$ with a
membership with a single positive entry assigned at random, we computed
$\vec{D}_{j\rightarrow\cdot}$, $\vec {E}_{\cdot\leftarrow k}$,
$Y_{gh}$, $n_{gh}$ accordingly from these initial values of $\vec{D}$
and $\vec{E}$, then initialized $p(D_{j\rightarrow k,g}=1,E_{j
\leftarrow k,h}=1)$ as detailed in the
supplement [\citet{Airoldi2013fk}]. 
The results of this experiment are shown in Table~\ref{estcomp-init}.

\begin{table}
\tabcolsep=0pt
\caption{Comparisons on $\varepsilon_B$ as the
estimation error of $B$ and the first highest membership accuracy
between different initialization for variational Bayes and MCMC}\label{estcomp-init}
\begin{tabular*}{\tablewidth}{@{\extracolsep{\fill}}lcc@{\hspace*{-3pt}}ccc@{\hspace*{-3pt}}c@{}}
\hline
& \multicolumn{3}{c}{\textbf{vEM}} & \multicolumn{3}{c@{}}{\textbf{MCMC}} \\[-6pt]
& \multicolumn{3}{c}{\hrulefill} & \multicolumn{3}{c@{}}{\hrulefill} \\
\textbf{Init.} & $\bolds{\varepsilon_B}$ & \textbf{Row} & \textbf{Column} & $\bolds{\varepsilon_B}$ & \textbf{Row} & \textbf{Column}\\
\hline
Random & 0.200 (0.163) & 0.916 (0.184) & 0.907 (0.120) & 0.171 (0.179) & 0.872 (0.171) &0.818 (0.177) \\
Uniform & 0.205 (0.173) & 0.916 (0.117) &0.880 (0.102) & 0.115 (0.175) & 0.957 (0.047) &0.820 (0.157) \\
\hline
\end{tabular*}
\end{table}

Another input for Algorithm \ref{algoVB} and the MCMC inference
algorithm is the Dirichlet parameters $\alpha$ and $\beta$. A priori,
$\alpha$, $\beta<1$ favor a single dominating membership component while
$\alpha$, $\beta>1$ favor diffuse membership.
In the analysis of real data, we expect few dominating memberships, so
we typically set $\alpha=\beta$ equal to either $0.2$ or $0.05$ and
assess sensitivity of resulting estimated memberships and other parameters.
However, the question arises as to whether an alternative strategy that
features informative priors is more useful than using noninformative as
we do. Using informative priors for the membership parameters might
lead to improved inference, especially in the case of substantial
nonidentifiability.

To evaluate this issue, we
generated a data set with informative priors $\vec{\alpha}=(0.3,
0.7)'$ for the rows and $\vec{\beta}=(0.6, 0.3, 0.10)'$ for the columns.
Then we fit the model with the vEM algorithm on this data set using both
noninformative uniform priors ($\alpha=\beta=0.05$) and
informative priors with the vectors $\vec{\alpha}, \vec{\beta}$ set
at the true values.
The results are presented in Table~\ref{estcomp-alpha} from which we
see that the results are comparable. This justifies the simple choice
of noninformative prior in our algorithms.

%
\begin{figure}

\includegraphics{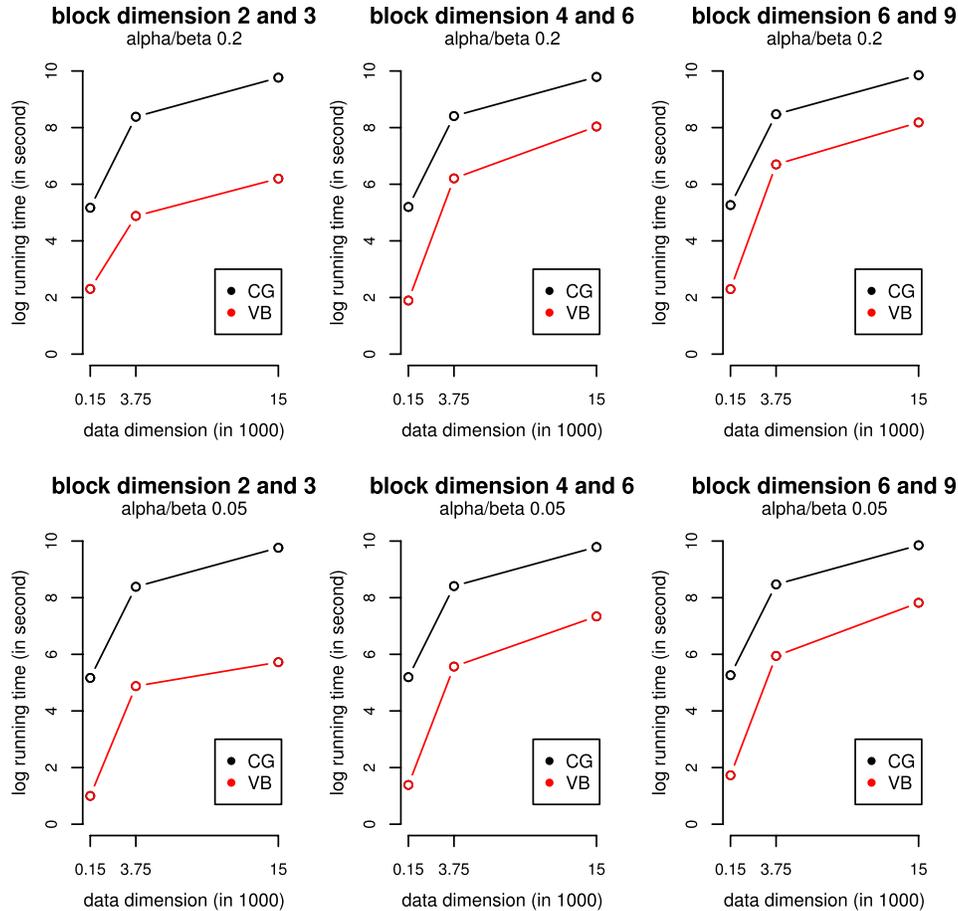}

\caption{Log run-time for simulated data. Red lines represent
variational Bayes and black lines represent MCMC via collapsed
Gibbs. The $x$-axis is the number of elements in a table. For
instance, 0.15 (thousand) represents a 10 by 15 table with 150 elements.} \label{figlogall-dBdecre}
\end{figure}



\begin{table}
\tabcolsep=0pt
\caption{Comparison of vEM fits using informative and
noninformative priors, in terms of estimation error $\varepsilon_B$ and
accuracy in estimating the highest membership component}\label{estcomp-alpha}
\begin{tabular*}{\tablewidth}{@{\extracolsep{\fill}}lc@{\hspace*{-10pt}}ccc@{\hspace*{-10pt}}c@{}}
\hline
\multicolumn{3}{@{}c}{\textbf{Noninformative priors}} & \multicolumn{3}{c@{}}{\textbf{Informative priors}} \\[-6pt]
\multicolumn{3}{@{}c}{\hrulefill} & \multicolumn{3}{c@{}}{\hrulefill} \\
$\bolds{\varepsilon_B}$ & \textbf{Row} &  \textbf{Column} & $\bolds{\varepsilon_B}$ & \textbf{Row} & \textbf{Column} \\
\hline
0.385 (0.176) & 0.868 (0.121) & 0.827 (0.134) & 0.203 (0.121) & 0.788 (0.157) &0.870 (0.091) \\
\hline
\end{tabular*}
\end{table}

\subsection{Run-time comparison}\label{secvb-mcmc}
As seen previously, variational Bayes performs as effectively as
MCMC in parameter and membership estimation as well as held-out
prediction accuracy. In the following, we present results on run-time
comparison between these two approaches. Our goal is to quantify
the magnitude of savings that variational Bayes can achieve while
obtaining similar inferences to those obtained through MCMC.


%

For each experiment we run 10 times, and the average log run-time is
recorded. The plots are shown in Figure~\ref{figlogall-dBdecre}. Three
table sizes are considered in this simulation: $10 \times15$, $50
\times75$ and $100 \times150$. From this figure, the run-time for MCMC
is consistently several times larger more than that of variational Bayes. For
example, when block sizes equal $(6, 9)$, and Dirichlet parameters equal
0.05, one experiment takes about 30 minutes to run for variational
Bayes, and it takes roughly 6~hours for MCMC. This trend continues when
table size increases, and the saving on computational cost can be much
more. These results suggest that variational Bayes should be preferred
for analyzing large tables. Recently developed inference strategies based on spectral clustering
[\citet{Rohe2012}] and binary factor graphs
[\citet{Azari2012}] should also be considered.

\subsection{Quantifying the effects of censoring}\label{secsimcensoring}
One of the issues in existing studies of coordinated cellular
responses is the preprocessing of the original
measurements. This kind of censoring reduces data utility and
decreases estimation \mbox{accuracy}. The goal of this study is to
quantify the effects of censoring by thresholding on the
estimation of the blockmodel.

The data $Y$ are generated from $Y(j,k) \sim \operatorname{Normal
}(\vec{\pi}_j' B\vec{p_k},\sigma^2)$. The domain of $Y(j,k)$ is
$(-\infty,+\infty)$. We perform Inverse Fisher Transformation (IFT)
that maps $Y(j,k)$ to $\rho(j,k)$ so that its range is $[-1,1]$. The
censored data are defined as $S(j,k)=\mathbf{1}(|\rho(j,k)|\geq\tau)$,
where $\tau$ can be median, mean or 0.5. Clearly, \mbox{$S(j,k) \in\{0,1\}$}.

The Normal blockmodel is applied to the original data $Y(j,k)$ and the
Bernoulli blockmodel described in Section~\ref{secngmodel} is applied
to the censored data $S(j,k)$. To make the comparison in the same
scale, we define $\hat{\rho}(j,k)$ as the IFT of
$\vec{\phi}_{j\rightarrow k}' \hat{B} \vec{\eta}_{j \leftarrow k}$,
where $\vec{\phi}_{j\rightarrow k}$, $\hat{B}$ and $\vec{\eta}_{j
\leftarrow k}$ are estimated from the Normal blockmodel. The estimation
error is defined as
\[
\varepsilon=\frac{\sum_{(j,k)}|\rho(j,k)-\hat{\rho}{(j,k)}|}{N_1
\times N_2}.
\]
The estimation error for the censored experiment is computed in
the same fashion, with $\hat{\rho}(j,k)=\vec{\phi}_{j\rightarrow
k}' \hat{B} \vec{\eta}_{j \leftarrow k}$, where
$\vec{\phi}_{j\rightarrow k}$, $\hat{B}$ and $\vec{\eta}_{j
\leftarrow k}$ are estimated from the Bernoulli blockmodel, and $\rho
(j,k)$ replaced by $|\rho(j,k)|$.



\begin{table}
\tabcolsep=0pt
\centering \caption{Comparison of estimation error on censored and
noncensored data. Standard errors are inside the~parenthesis}\label{censormodel61e}
\begin{tabular*}{\tablewidth}{@{\extracolsep{\fill}}lccd{1.10}d{1.10}d{1.10}@{\hspace*{-3pt}}}
\hline
\textbf{Data} & \textbf{Method} & \textbf{Bic.}
& \multicolumn{1}{c}{\textbf{Error} $\bolds{\varepsilon}$} & \multicolumn{1}{c}{\textbf{Recall}} & \multicolumn{1}{c}{\textbf{Precision}} \\
\hline
Raw $\rho_{ij}$ & 2-way Normal & 6 & 0.054\mbox{ }(0.010) & 0.841\mbox{ }(0.169) & 0.881\mbox{ }(0.116) \\
& Hier. clustering & 6 & 0.221\mbox{ (--)} & 0.967\mbox{ (--)} & 0.970\mbox{ (--)} \\
& Cheng \& Church & 2 & \multicolumn{1}{c}{--} & 0.367\mbox{ (--)} & 0.679\mbox{ (--)}
\\[3pt]
$|\rho_{ij}| > \rho_{(0.5)}$ & 2-way Bernoulli & 6 & 0.175\mbox{ }(0.006) & 0.518\mbox{ }(0.017) & 0.722\mbox{ }(0.048) \\
& Hier. clustering & 6 & 0.125\mbox{ (--)} & 0.700\mbox{ (--)} & 0.850\mbox{ (--)} \\
& Cheng \& Church & 2 & \multicolumn{1}{c}{--} & 0.232\mbox{ (--)} & 0.640\mbox{ (--)} \\
& Cross-associations & 4 & \multicolumn{1}{c}{--} & 0.667\mbox{ (--)} & 0.762\mbox{ (--)}
\\[3pt]
$|\rho_{ij}| > \bar\rho$ & 2-way Bernoulli & 6 & 0.182\mbox{ }(0.003) & 0.528\mbox{ }(0.014) & 0.773\mbox{ }(0.056) \\
& Hier. clustering & 6 & 0.187\mbox{ (--)} & 0.500\mbox{ (--)} & 0.841\mbox{ (--)} \\
& Cheng \& Church & 3 & \multicolumn{1}{c}{--} & 0.237\mbox{ (--)} & 0.640\mbox{ (--)} \\
& Cross-associations & 6 & \multicolumn{1}{c}{--} & 0.667\mbox{ (--)} & 0.841\mbox{ (--)}
\\[3pt]
$|\rho_{ij}| > 0.5$ & 2-way Bernoulli & 6 & 0.158\mbox{ }(0.002) & 0.528\mbox{ }(0.022) & 0.835\mbox{ }(0.030) \\
& Hier. clustering & 6 & 0.189\mbox{ (--)} & 0.500\mbox{ (--)} & 0.841\mbox{ (--)} \\
& Cheng \& Church & 3 & \multicolumn{1}{c}{--} & 0.239\mbox{ (--)} & 0.640\mbox{ (--)}\\
& Cross-associations & 8 & \multicolumn{1}{c}{--} & 0.613\mbox{ (--)} & 0.667\mbox{ (--)} \\
\hline
\end{tabular*}
\end{table}

We compare our model with a bi-clustering method popular in
computational biology [\citet{cheng2000biclustering}], fit to both
the raw and censored correlations.
We match each estimated bicluster to a true block and compute recall
and precision in estimating absolute correlations above a threshold.
Results are presented in Table~\ref{censormodel61e}, where the results
obtained with BCCC are optimized over a range of input parameter
values. For completeness, we also add results obtained with
hierarchical clustering to rows and columns independently, and with
cross-association [\citet{ChakPapaModhFalo2004}].

The effects of censoring are clearly seen from
Table~\ref{censormodel61e}. The estimation error increases more than
threefold when using the censored data with the Bernoulli block model.
The effect of thresholding parameter $\tau$ is not very significant.


\section{Analyzing transcriptional and metabolic coordination in
response to starvation} \label{secmet}

Functions in a cell are executed by cascades of molecular events.
Intuitively, proteins are the messengers, while metabolites and other
small molecules are the messages. Measuring protein activity over time,
directly, is difficult and expensive. An indication of the abundance of
most proteins, however, can be inferred from the amount of the
messenger RNA transcripts. These transcripts are copies of genes and
lead to the translation of proteins. This is especially true in yeast
where alternatives to the transcription-translation hypothesis, such as
alternative splicing, are not frequent. Metabolite concentrations add
an essential perspective to the study of cascades of molecular events.

We conducted an integrated analysis of two data collections recently
published: temporal profiles of metabolite concentrations
[\citet{brauyuanbennlu2006}] and temporal profiles of gene
expression [\citet{bradbraurabitroy2009}], both measured in
\textit{Saccharomyces cerevisiae} with matching sampling schemes.

An integrated analysis of the coordination between gene expression
and metabolite concentrations may lead to the identification of
sets of genes (i.e., the corresponding proteins) and metabolites
that are functionally related, which will provide additional
insights into regulatory mechanisms at multiple levels and open
avenues of inquiry. The identification and quantification of such
coordinated regulatory behavior is the goal of our analysis.

The methodology in Section~\ref{secmix-pq} allows us to identify genes
and metabolites that show correlated responses to metabolic stress,
namely, starvation. To evaluate the biological significance of the
results, we quantify to what extent correlated responses are associated
with metabolic-related functions and to what extent estimated models
can be used to identify functionally related genes and metabolites
out-of-sample.\newpage
\subsection{Data and experimental design}\label{secdata}
The expression data consist of messenger RNA transcript levels measured
using Agilent microarrays on cultures of \textit{S. cerevisiae} before
and after carbon starvation (glucose removal), and before and after
nitrogen starvation (ammonium removal). Collection times were 0 minutes
(before starvation) and 10, 30, 60, 120, 240 and 480 minutes after
starvation. For more details about the data and the experimental
protocol see \citet{bradbraurabitroy2009}.
The metabolite concentrations data were obtained using liquid
chromatography-mass spectrometry before and after carbon starvation
(glucose removal), and before and after nitrogen starvation (ammonium
removal). Collection times were 0 minutes (before starvation) and 10,
30, 60, 120, 240 and 480 minutes after starvation. For more details
about the data and the experimental protocol see
\citet{brauyuanbennlu2006}.

The concentration of each metabolite and the transcript level of each
gene at time point $t$ are expressed as $\log_2$ ratios versus the
corresponding measurements at the zero time point. Thus, for each gene
$j$ we have a sequence $G_{jt}$, $t=1,\ldots,6$, and for each
metabolite $k$ we have a sequence $M_{kt}$, $t=1,\ldots,6$,
representing for the 6 time points observation after time 0. Complete
temporal profiles are available for 5039 genes and for 61 metabolites;
783 genes and 7 metabolites with missing data were not considered.

Using the temporal profile, we can calculate the sample
correlation coefficient of each gene and metabolite pair $(j,k)$:
$\rho(j,k) = \frac{\sum_{t=1}^T (G_{jt}-\bar G_{j}) (M_{kt}-\bar
M_{k})}{(T-1)S_GS_M}$, where $\bar G_{j} = \sum_t G_{jt}/T$ and
$\bar M_{k} = \sum_t M_{kt}/T$ are the sample mean, and $S_G$ and
$S_M$ are the sample standard deviation. We then transform these
correlations using the Fisher transformation
$Z(j,k)=\frac{1}{2}\log\frac{1+\rho{(j,k)}}{1-\rho{(j,k)}}$.
With the true correlation between genes and metabolites denoted as
$\rho_0$, we have $Z(j,k)$ following asymptotically Normal distribution
with mean $\mu_z = \frac{1}{2}\log\frac{1+\rho_0}{1-\rho_0}$ and
standard error $1/\sqrt{T-3}$ [\citeauthor{Fish1915}
(\citeyear{Fish1915,Fish1921})]. Under the hypothesis that there is no
correlation between genes and metabolites, we will expect $\rho_0=0$
and $\mu_z = \frac{1}{2}\log\frac{1+\rho_0}{1-\rho_0}=0$.
These Fisher transformed quantities provide the input to our model.

We do not expect the multi-way blockmodel assumption to hold for all
the 5039 genes. Instead, we provide separate joint analyses on subsets
of genes and all 61 metabolites, for a number of gene lists of
interest, which we expect to be involved in the cellular response to
starvation. We consider gene lists that were obtained in studies
exploring the environmental stress response (ESR), cellular
proliferation, metabolism and the cell cycle
[\citet{GascSpelKaoCarm2000}, \citet{TuKudlRowiMcKn2005},
\citet{BrauHuttAiroRose2007}, \citeauthor{AiroHuttGresLu2008} (\citeyear{AiroHuttGresLu2008,Airoldi2013uq}),
\citet{Slavov2013fk}].

For all the experiments we rely on variational Bayes implementation of
our model due to its advantage in convergence speed, which is crucial
when dealing with correlation tables involving hundreds of genes. We
adopt the setting as described in Section~\ref{secvb-mcmc} for VB, with
specific changes described as they become relevant. In the remainder of
this section, with the exception of Section~\ref{sec4.5}, we
consistently set the number of metabolite blocks $K_2=4$, since there
are four metabolite classes, and we use informative priors for the
memberships of each metabolite, depending on which class they are known
to belong to. Specifically, each of the 61 metabolites belongs to one
of the four classes: TCA, AA, GLY, BSI. If a metabolite is in class
TCA, say, Aconitate, its $\vec{\xi}$ vector will be initialized as
$\vec{\xi} ={}$[100 1 1 1], normalized to unit norm. By assuming a
dominating component on the true index in the initial membership, the
metabolites will mostly remain stably associated to their classes
during VB inference.

For the optimal number of gene blocks, we select $K_1$ by minimizing
the Bayesian information criterion (BIC). The
general BIC formula is $-2\log L+k\times\log(n)$, in which $k$ is
the number of parameters, $n$ is the number of observations, and
$L$ is the likelihood. For our model, the approximated BIC is
\[
-2\log L+|B,\vec{\pi},\vec{p}| \times\log|Y|,
\]
where $|B,\vec{\pi},\vec{p}|$ is the number of parameters, which
is approximately equal to \mbox{$K_1\times K_2$,} and $|Y|$
is the number of entries in the table, that is, $|Y|=N_1\times N_2$.

In Section~\ref{secmet} we present some of the results with the goal of
showcasing how the data analysis, via the multi-way block model,
supports the biological research.


\subsection{Multifaceted functional evaluation of coordinated responses}\label{secgo-eval}
Here we evaluate to what extent the proposed model is useful in
revealing the genes' multifaceted functional roles. We rely on the
functional enrichment analysis using the Gene Ontology to evaluate the
functional content of clusters of genes [\citet{Ashbetal2000},
\citet{BoylWengGollJin2004}].

\begin{table}[b]
\tabcolsep=0pt
\caption{Example functional evaluation.
Gene Ontology terms associated with first- and second-largest
membership scores for the Nitrogen starvation experiment}\label{tabfunctionalevaluationnit}
%
\begin{tabular*}{\tablewidth}{@{\extracolsep{\fill}}lcccd{1.7}@{}}
\hline
\textbf{Memb.} & \textbf{Class} & \textbf{Ontology} & \textbf{Term description} & \multicolumn{1}{c@{}}{\textbf{$\bolds{p}$-value}} \\
\hline
First & AA & Component & DNA-directed RNA polymerase I complex & 9.8\mbox{E--}6\\
First & AA & Component & Preribosome, small subunit precursor & 0.00324\\
First & AA & Function & Translation factor activity, nucleic acid binding & 5.11\mbox{E--}14 \\
First & AA & Function & Translation initiation factor activity & 1.64\mbox{E--}10 \\
First & AA & Function & DNA-directed RNA polymerase activity & 1.45\mbox{E--}6\\
First & BSI & Component & Preribosome, large subunit precursor & 0.00013 \\
First & BSI & Function & GTP binding & 0.03314 \\
First & BSI & Function & Guanyl ribonucleotide binding & 0.03314
\\[3pt]
Second & AA & Component & DNA-directed RNA polymerase III complex & 6.76\mbox{E--}10 \\
Second & AA & Component & DNA-directed RNA polymerase II core complex & 0.00322 \\
Second & AA & Function & RNA polymerase activity & 1.27\mbox{E--}6 \\
Second & AA & Function & ATP-dependent RNA helicase activity & 3.43\mbox{E--}6\\
Second & BSI & Component & Ribonucleoprotein complex & 6.47\mbox{E--}12 \\
Second & BSI & Component & 90S preribosome & 0.00124 \\
Second & BSI & Function & Aminoacyl-tRNA ligase activity & 0.0000817\\
Second & BSI & Function & N-methyltransferase activity & 0.0455 \\
\hline
\end{tabular*}
\end{table}

One aspect of our model that distinguishes it from clustering and
bi-clustering methods is the mixed membership assumption. That is,
in our model, each gene can participate in multiple functions, as
modeled via the gene-specific latent membership vectors
$\vec{\pi}$. In practice, the membership assumption lets us
identify multiple levels of functional enrichment.

To illustrate this point, we consider 521 genes that were found to be
strongly associated with metabolic activities, that is, up-regulated in
response to increasing growth rate, in previous studies
[\citet{BrauHuttAiroRose2007}, \citet{AiroHuttGresLu2008}].
%
We use the largest estimated memberships for each gene $\pi_{gi}$, $i=1,
\ldots,K_1$, to assign genes $g=1,\ldots,521$ to metabolite classes
$j=1,\ldots,4$. Then we perform functional analysis on the resulting
sets of genes associated with each metabolite class.

More formally, we proceed as follows. First, the largest estimated
membership is used to assign gene $g$ to gene block $i$, according to
$\hat i_g =\break  \arg\max_{i=1,\ldots,K_1} \pi_{gi}$.
Then the largest estimated gene-block to metabolite-class association
$|B_{ij}|$ is then used to assign gene $g$ with a metabolite class,
according to
\[
\hat j_g = \mathop{\arg\max}_{j=1,\ldots,4}
|B_{\hat{i}_g,j}|.
\]
The collection of estimated gene-to-metabolite class associations, $\{
\hat j_g$, $g=1,\ldots,521\}$, is used to partition genes into four sets,
for example, $AA = \{g \mbox{ s.t. } \hat j_g=1\}$. We perform
functional enrichment analysis for each of these four sets.
In addition, the mixed membership nature of the proposed multi-way
blockmodel allows us to analyze second-order functional
enrichment. We repeat the procedure above but we estimate $\hat
i_g$ using the second-largest membership in $\vec\pi_g$.

The functional analysis results obtained for both first- and
second-largest memberships are reported in
Table~\ref{tabfunctionalevaluationnit}.
Interestingly, subsets of genes associated with the
same metabolite class, for instance, AA, are functionally enriched
for multiple functions, to different degrees. For instance, genes
use AA metabolites when performing translational activities in the
nucleus primarily, however, they use AA metabolites when
performing polymerase-related activities on the polymerase
II~and~II complexes to a lesser extent. Similarly, genes use BSI
metabolites for binding activities in the preribosome primarily,
and for ligase and transferase activities in the preribosome and
the ribonucleoprotein complex\vadjust{\goodbreak} to a minor extent. The magnitude of
the components of the relevant mixed membership vectors provides
more information on the degree of involvement the various gene
blocks in these many activities. This type of multifaceted
functional analysis is possible thanks to the mixed membership
assumption encoded in the multi-way blockmodel.

These results highlight the role of the mixed membership assumption in
supporting a detailed multifaceted functional analysis, which is not
possible with traditional methods.


\subsection{Predicting functional annotations out-of-sample}\label{secrealoos}
Here we assess the goodness of fit of the proposed method on real
data, in terms of predictions out-of-sample. We present results of
an experiment
in which we predict held-out functional annotations $\pi_{gi}$.
This analysis leverages use of informative priors on a subset of known
functional annotations.

We consider 57 genes that were found in previous studies to be strongly
associated with cellular growth
[\citet{BrauHuttAiroRose2007,AiroHuttGresLu2008}], 760 genes that
were found to be involved in the environmental response to stress
[\citet{GascSpelKaoCarm2000}], and 19 genes that were found to be
involved in metabolic cycling [\citet{TuKudlRowiMcKn2005}].
Good out-of-sample prediction performance will enable biologists to use
this method to guide which functions they should be testing at the
bench, speeding up the exploration of the functional landscape through
statistical analysis of gene--metabolite associations.
%

\begin{table}[b]
\tabcolsep=0pt
\caption{Statistics for the lists of genes. Column
three reports the number of genes with one, some and no functional
annotations. $K_1$ is the number of gene blocks in the fitted blockmodel} \label{prior-stat}
\begin{tabular*}{\tablewidth}{@{\extracolsep{\fill}}ld{3.0}ccd{2.2}d{2.1}d{2.0}d{2.0}d{2.2}c@{}}
\hline
& \multicolumn{2}{c}{\textbf{No. of genes}} & \multicolumn{6}{c@{}}{\textbf{No. of functional annotations}} \\[-6pt]
& \multicolumn{2}{c}{\hrulefill} & \multicolumn{6}{c@{}}{\hrulefill} \\
\textbf{Gene list} & \multicolumn{1}{c}{\textbf{Total}} & \multicolumn{1}{c}{\textbf{One/some/none}}
                   & \multicolumn{1}{c}{\textbf{Min}} & \multicolumn{1}{c}{\textbf{25\%}}
                   & \multicolumn{1}{c}{\textbf{50\%}} & \multicolumn{1}{c}{\textbf{75\%}}
                   & \multicolumn{1}{c}{\textbf{Max}} & \multicolumn{1}{c}{\textbf{Mean}}
                   & $\bolds{K_1}$\\
\hline
Growth rate & 57 & $5/19/38$ & 1 & 1.25 & 4 & 7 & 7 & \phantom{0}4.26 & 12\\
ESR induced & 240 & $0/215/25$ & 2 & 5 & 7 & 12 & 31 & 9.31 & 76\\
ESR repressed & 520 & $1/503/17$ & 1 & 10 & 19 & 22 & 31 & 16.93 & 78\\
Metabolic cycle & 19 & $4/14/5$ & 1 & 1 & 6.5 & 10 & 20 & 7.29 & 25\\
\hline
\end{tabular*}
\end{table}

To establish the ground truth for this experiment, we collected
functional annotations for each gene in the same four lists as in
Section~\ref{secgo-eval} which will be held-out and predicted using the
multi-way blockmodel.
Table~\ref{prior-stat} reports summary statistics of the functional
annotations in each list of genes, obtained using the Gene Ontology
term finder (\citeauthor{SGD}). Column two reports the total number of genes in
each list. Column three reports the number of genes with one, some and
no functional annotations. Columns 4--9 report the quantiles from the
distribution of the number of functional annotations for the genes in
each list. Column 10 reports the value of~$K_1$ we selected for fitting
the blockmodel.\vadjust{\goodbreak}

\begin{table}[b]
\tabcolsep=10pt
\caption{Out-of-sample predictions of
functional annotations for the Nitrogen starvation experiment. Accuracy
in recovering (single/multiple) annotations for four lists of genes}\label{tabrealoosfunction}
\begin{tabular*}{\textwidth}{@{\extracolsep{\fill}}@{}lcccccccc@{}}%
\hline
& \multicolumn{4}{c}{\textbf{Single annotations}} & \multicolumn{4}{c@{}}{\textbf{Multiple annotations}} \\[-6pt]
& \multicolumn{4}{c}{\hrulefill} & \multicolumn{4}{c@{}}{\hrulefill} \\
& \multicolumn{2}{c}{\textbf{Observed}} & \multicolumn{2}{c}{\textbf{Missing}} & \multicolumn{2}{c}{\textbf{Observed}} & \multicolumn{2}{c@{}}{\textbf{Missing}} \\[-6pt]
& \multicolumn{2}{c}{\hrulefill} & \multicolumn{2}{c}{\hrulefill} & \multicolumn{2}{c}{\hrulefill} & \multicolumn{2}{c@{}}{\hrulefill} \\
\textbf{Gene list} & \textbf{0s} & \textbf{1s} & \textbf{0s} & \textbf{1s} & \textbf{0s} & \textbf{1s} & \textbf{0s} & \textbf{1s} \\
\hline
Growth rate & 0.94 & 0.36 & 0.94 & 0.36 & 0.83 & 0.75 & 0.84 & 0.74 \\
ESR induced & -- & -- & -- & -- & 0.92 & 0.39 & 0.92 & 0.46 \\
ESR repressed & -- & -- & 0.99 & 0.00 & 0.84 & 0.43 & 0.85 & 0.50 \\
Metabolic cycle & 0.97 & 0.25 & 0.96 & 0.10 & 0.78 & 0.65 & 0.80 & 0.63\\
\hline
\end{tabular*}
\end{table}

To perform the second experiment, we held out the annotations for 50\%
of the genes with multiple functional annotations, and we also held out
the annotation for 50\% of the genes with a single functional
annotation. When fitting the multi-way blockmodel, in addition to using
informative priors for the memberships of each metabolite depending on
which class they are known to belong, as detailed in
Section~\ref{secdata}, we used informative priors for the functional
annotations we \textit{did not} hold out. For the held-out annotations,
we used noninformative values for the hyperparameters instead.
For instance, suppose that the known vector of functional annotations
for gene $g$ is $\vec a_g ={}$[1 0 0 1 1 0 0 0 0 1], and that $a_g(1)$
and $a_g(4)$ were to be held out in a particular replication, so that
we have $\vec a_g ={}$[NA 0 0 NA 1 0 0 0 0 1]. The prior for the
functional annotation for that gene would be set at $\vec{\xi}_g ={}$[1
1 1 1 100 1 1 1 1 100], normalized to unit norm.
The rationale for this choice is to fit a multi-way blockmodel
with known biological structure for those genes and metabolites
that are used for parameter estimation, but agnostic about the
biology we want to predict out-of-sample.
We claimed success in each prediction if the imputed annotations, $\hat
\xi_g(k)=1$, corresponded to real held-out annotations, and if the
imputed absences of annotations, $\hat\xi_g(i)=0$, corresponded to
absences of real held-out annotations.
We repeated this procedure 10 times, for each of the four lists of genes.

Table~\ref{tabrealoosfunction} reports the accuracy results, detailed
by genes with single and multiple annotations, and evaluated separately
for annotations (i.e., the 1s) and lack of annotations (i.e., the 0s).
The baseline accuracy for predicting single annotations, using random
guesses for each gene independently, ranges between $1/19\approx5\%$
for the metabolic cycle genes to $1/520\approx0.2\%$ for the ERS genes.
The baseline accuracy is slightly higher for predicting multiple
annotations, since predicting single annotations is a harder problem.

For completeness, we also report the accuracy in predicting annotations
that were known during model fitting to get a sense of the goodness of
fit from a substantive, biological perspective. In fact, if the model
assumptions are accurate, we would expect accurate predictions for the
known annotations. If the model is not accurate, or if the model
provides too much shrinkage, we would expect lower accuracy on known
annotations.

Overall, the blockmodel assumptions are substantiated by the results in
Table~\ref{tabrealoosfunction}. The model is useful for encoding
biological information about single and multiple functional
annotations.
The out-of-sample prediction accuracy of the multi-way blockmodel is
solid and consistently much higher than the baseline. These results
complement and confirm the out-of-sample prediction results we obtained
in Section~\ref{secsimoos}.


\subsection{Coordinated and differential regulatory response to Nitrogen and Carbon starvation}
Here we provide an illustration of how the multi-way blockmodel
can be used to perform quantitative and qualitative analysis of
coordinated regulation in response to Nitrogen starvation and
differential regulation in response to Carbon starvation.
We perform this analysis for the same four lists of genes we considered
in Section~\ref{secgo-eval}.

\begin{table}[b]
\tabcolsep=10pt
\caption{Quantitative evaluation of
coordinated regulatory responses. Number of genes associated with the
same metabolite class in both the Nitrogen and Carbon starvation
experiments. The association is estimated using both largest and
second-largest membership scores}\label{tabsameassoc}
\begin{tabular*}{\textwidth}{@{\extracolsep{\fill}}@{}ld{3.0}d{2.0}d{2.0}d{2.0}d{3.0}d{2.0}cd{2.0}@{}}%
\hline
& \multicolumn{4}{c}{\textbf{Largest membership}} & \multicolumn{4}{c@{}}{\textbf{Second-largest membership}} \\[-6pt]
& \multicolumn{4}{c}{\hrulefill} & \multicolumn{4}{c@{}}{\hrulefill} \\
\textbf{Gene list} & \multicolumn{1}{c}{\textbf{AA}} & \multicolumn{1}{c}{\textbf{BSI}} & \multicolumn{1}{c}{\textbf{GLY}} & \multicolumn{1}{c}{\textbf{TCA}}
                   & \multicolumn{1}{c}{\textbf{AA}} & \multicolumn{1}{c}{\textbf{BSI}} & \multicolumn{1}{c}{\textbf{GLY}} & \multicolumn{1}{c@{}}{\textbf{TCA}} \\
\hline
Growth rate & 11 & 6 & 10 & 2 & 10 & 5 & 3 & 1 \\
ESR induced & 55 & 13 & 4 & 0 & 48 & 21 & 4 & 0 \\
ESR repressed & 128 & 22 & 1 & 17 & 109 & 52 & 0 & 27 \\
Metabolic cycle & 2 & 3 & 1 & 0 & 2 & 3 & 1 & 0 \\
\hline
\end{tabular*}
\end{table}

The quantitative analysis of \textit{coordinated regulation} is based
on the number of genes which are estimated to be associated with the
various metabolite classes, in both the Nitrogen and the Carbon
starvation experiments.

We used the same procedure described in Section~\ref{secgo-eval} to
estimate the metabolite classes associated with each gene, using the
estimated largest and second-largest (gene-block) memberships.
Table~\ref{tabsameassoc} reports the number of genes that were found to
be associated with a primary metabolite class (largest membership) and
with a~secondary metabolite class (second-largest membership) for each
of the four lists of genes we consider.

About a fourth of the genes are found to be associated with a
primary metabolite class.
Despite the similarity in the patterns of primary and secondary
associations, the gene sets involved in them are different. These
results imply that another fourth of the
genes are found to be associated to a secondary metabolite class.
Overall, the blockmodel suggests a substantial amount of overlap
between the coordinated regulatory response to Nitrogen and Carbon
starvation. A similar quantitative analysis could be conducted for highlighting
Nitrogen- and Carbon-specific coordinated regulatory responses.
%
\begin{table}[b]
\tabcolsep=0pt
\caption{Functional evaluation of
gene--metabolite associations that are differentially regulated in
Nitrogen and Carbon. Gene Ontology terms for gene--metabolite
associations unique to the Nitrogen starvation experiment. Association
is computed using the largest memberships}\label{tabfunctionalevaluationdiff}
\begin{tabular*}{\tablewidth}{@{\extracolsep{\fill}}@{}lccd{1.7}@{}}
\hline
\textbf{Class} & \textbf{Ontology} & \textbf{Term description} & \multicolumn{1}{c}{$\bolds{p}$\textbf{-value}} \\
\hline
AA & Function & Alcohol dehydrogenase (NADP$+$) activity & 0.01775 \\
AA & Function & Aldo-keto reductase (NADP) activity & 0.01775 \\
AA & Process & Vacuolar protein catabolic process & 0.00026 \\
AA & Process & Catabolic process & 0.00808 \\
BSI & Function & Peroxidase activity & 0.0000568 \\
BSI & Function & Antioxidant activity & 0.0004 \\
BSI & Function & Carbohydrate kinase activity & 0.00083 \\
BSI & Function & Glutathione peroxidase activity & 0.0214 \\
BSI & Process & Carbohydrate catabolic process & 2.98\mbox{E--}7 \\
BSI & Process & Cellular response to oxidative stress & 0.0000131 \\
BSI & Process & Trehalose metabolic process & 0.0000203 \\
BSI & Process & Alcohol catabolic process & 0.0000231 \\
BSI & Process & Glycoside metabolic process & 0.000061 \\
GLY & Function & Oxidoreductase activity & 0.0000116 \\
GLY & Process & Oxidation--reduction process & 0.00097 \\
GLY & Process & Cellular carbohydrate metabolic process & 0.0011 \\
GLY & Process & Carbohydrate metabolic process & 0.00143 \\
GLY & Process & Cellular aldehyde metabolic process & 0.00351 \\
\hline
\end{tabular*}
\end{table}

The qualitative analysis of \textit{differential regulation} is based
on the functional enrichment analysis of those genes associated with a
given metabolite class in the Nitrogen experiment, but associated with
a different metabolic class in the Carbon experiment.
For this analysis, we used the procedure above to estimate the
metabolite classes associated with each gene, using the estimated
largest memberships only, for the list of genes that were found to be
ESR induced.
The results of the functional analysis obtained for the largest
memberships, using the Gene Ontology term finder, are reported in
Table~\ref{tabfunctionalevaluationdiff}.

These results highlight how the same set of genes (a proxy for
proteins) may be using metabolites differently to execute a response to
the Carbon. For instance, metabolites in the BSI class are used to
process glycoside, alcohol and trehalose, as part of antioxidant
activities. Metabolites in the GLY class are used to execute
oxidoreductase activities and metabolize aldehyde and carbohydrates.
The magnitude of the components of the relevant mixed membership
vectors provides more information on the degree of involvement the
various gene blocks in these many activities.

A similar qualitative analysis could be carried out to explore the
functional landscape, that is, shared by the Nitrogen and Carbon
coordinated regulatory responses to starvation.


\subsection{Comparative analysis of raw and preprocessed data}\label{sec4.5}
Here we compare a blockmodel analysis of coordinated regulation with an
analysis using cross-association \citet{ChakPapaModhFalo2004},
quantitatively, in terms of number of gene--metabolite class
associations found.
We consider the four lists of genes above for this analysis.

Cross-association takes a binary table as input. We built such a
genes-by-metabolites binary matrix $Y$ by thresholding the
corresponding matrix of correlations.
We assign $Y(j,k) = 1$ whenever $\rho(j,k)$ is above the 75th
percentile or below the 25th percentile of the empirical
correlation distribution.


Cross-association provides a two-way blockmodel as output, in
which $K_1$ and $K_2$ are estimated using a metric based on
information gain.
To make a valid comparison, we fit the stochastic
multi-way blockmodel with the same number of gene and metabolite
blocks.

An additional complication in this analysis is that the number of
metabolite blocks can be different from four, for both
cross-association and the stochastic blockmodel. We use
noninformative priors on the metabolites memberships in the
stochastic blockmodel. In addition, we developed a greedy
matching procedure to associate metabolite blocks to metabolite
classes, after inference. We proceeded as follows.
Each metabolite was associated with a block using its largest
(metabolite-block) membership.
Each metabolite is associated with a known metabolite class.
We assigned a metabolite class label to each metabolite block according
to a simple majority rule.

We used the same procedure described in Section~\ref{secgo-eval} to
estimate the metabolite classes associated with each gene, using the
estimated largest and second-largest (gene-block) memberships.

Table~\ref{tabcompcrossassoc} reports the number of Gene Ontology terms
that were found to be associated with a primary metabolite class
(largest membership) in the first four rows, and with a secondary
metabolite class (second-largest membership) in the next four rows, for
each of the four lists of Gene Ontology terms we consider. The last
four rows report the number of genes that were found to be associated
with a metabolite class using cross-association.
The multi-way stochastic blockmodel finds more primary
associations than cross-association, 246 versus 194. In addition,
if we consider the secondary associations, the blockmodel
analysis uncovers 190 more associations.
In fact, subsets of genes associated with the same primary and
secondary metabolite class, for instance, AA, are not overlapping by
construction.

\begin{table}
\tabcolsep=0pt
\caption{Quantitative evaluation of
Gene Ontology terms associated with gene--metabolite class found. Shown
in the tables are results for multi-way blockmodel's largest (1st) and
second-largest (2nd) memberships as well as cross-associations (CA)}\label{tabcompcrossassoc}
%
\begin{tabular*}{\tablewidth}{@{\extracolsep{\fill}}@{}lcd{2.0}d{2.0}d{2.0}d{2.0}d{2.0}d{2.0}cccd{3.0}@{}}
\hline
& & \multicolumn{3}{c}{\textbf{AA}} & \multicolumn{3}{c}{\textbf{BSI}} & \multicolumn{3}{c}{\textbf{TCA}} & \\[-6pt]
& & \multicolumn{3}{c}{\hrulefill} & \multicolumn{3}{c}{\hrulefill} & \multicolumn{3}{c}{\hrulefill} & \\
\textbf{Memb.} & \textbf{Gene list} &
\multicolumn{1}{c}{\textbf{BP}} & \multicolumn{1}{c}{\textbf{CC}} & \multicolumn{1}{c}{\textbf{MF}} &
\multicolumn{1}{c}{\textbf{BP}} & \multicolumn{1}{c}{\textbf{CC}} & \multicolumn{1}{c}{\textbf{MF}} &
\multicolumn{1}{c}{\textbf{BP}} & \multicolumn{1}{c}{\textbf{CC}} & \multicolumn{1}{c}{\textbf{MF}} & \multicolumn{1}{c@{}}{\textbf{Total}} \\
\hline
1st & Growth rate & 1 & \multicolumn{1}{c}{--} & \multicolumn{1}{c}{--} & 1 & 6 & 5 & \multicolumn{1}{c}{--} & \multicolumn{1}{c}{--} & \multicolumn{1}{c}{--} & 13 \\
1st & ESR induced & 33 & 7 & 11 & 2 & 1 & 2 & 16 & \multicolumn{1}{c}{--} & 5 & 77 \\
1st & ESR repressed & \multicolumn{1}{c}{--} & 45 & 26 & 32 & 23 & 5 & \multicolumn{1}{c}{--} & \multicolumn{1}{c}{--} & \multicolumn{1}{c}{--} & 131 \\
1st & Metabolic cycle & 13 & 6 & 6 & \multicolumn{1}{c}{--} & \multicolumn{1}{c}{--} & \multicolumn{1}{c}{--} & \multicolumn{1}{c}{--} & \multicolumn{1}{c}{--} & \multicolumn{1}{c}{--} & 25
\\[3pt]
2nd & Growth rate & 4 & 6 & 3 & \multicolumn{1}{c}{--} & \multicolumn{1}{c}{--} & \multicolumn{1}{c}{--} & \multicolumn{1}{c}{--} & \multicolumn{1}{c}{--} & \multicolumn{1}{c}{--} & 13 \\
2nd & ESR induced & \multicolumn{1}{c}{--} & 1 & 6 & \multicolumn{1}{c}{--} & 16 & 14 & \multicolumn{1}{c}{--} & \multicolumn{1}{c}{--} & 1 & 38 \\
2nd & ESR repressed & \multicolumn{1}{c}{--} & 47 & 21 & \multicolumn{1}{c}{--} & 34 & 11 & \multicolumn{1}{c}{--} & \multicolumn{1}{c}{--} & \multicolumn{1}{c}{--} & 113 \\
2nd & Metabolic cycle & \multicolumn{1}{c}{--} & \multicolumn{1}{c}{--} & \multicolumn{1}{c}{--} & 12 & 6 & 8 & \multicolumn{1}{c}{--} & \multicolumn{1}{c}{--} & \multicolumn{1}{c}{--} & 26
\\[3pt]
CA & Growth rate & \multicolumn{1}{c}{--} & 6 & 2 & 2 & \multicolumn{1}{c}{--} & 2 & \multicolumn{1}{c}{--} & \multicolumn{1}{c}{--} & \multicolumn{1}{c}{--} & 12 \\
CA & ESR induced & \multicolumn{1}{c}{--} & \multicolumn{1}{c}{--} & \multicolumn{1}{c}{--} & \multicolumn{1}{c}{--} & 21 & 19 & \multicolumn{1}{c}{--} & \multicolumn{1}{c}{--} & \multicolumn{1}{c}{--} & 40 \\
CA & ESR repressed & \multicolumn{1}{c}{--} & 47 & 20 & \multicolumn{1}{c}{--} & 30 & 20 & \multicolumn{1}{c}{--} & \multicolumn{1}{c}{--} & \multicolumn{1}{c}{--} & 117 \\
CA & Metabolic cycle & \multicolumn{1}{c}{--} & \multicolumn{1}{c}{--} & \multicolumn{1}{c}{--} & 12 & 6 & 7 & \multicolumn{1}{c}{--} & \multicolumn{1}{c}{--} & \multicolumn{1}{c}{--} & 25 \\
\hline
\end{tabular*}
\end{table}

Overall, cross-association is not well suited for any analysis of
biological correlations because of a number of shortcomings,
including its reliance on binary input and its lack of
flexibility for incorporating prior biological information, for example,
the number of metabolite blocks. Our results show that the
multi-way stochastic blockmodel outperforms cross-associations
quantitatively, even when we do not make use of biological prior
knowledge.


\section{Concluding remarks}
\label{secsummary}

In order to analyze the temporal coordination between gene expression
and metabolite concentrations in yeast cells, in response to
starvation, we developed a family of multi-way stochastic blockmodels.
These models extend the mixed membership stochastic blockmodel
[\citet{AiroBleiFienXing2008}] to the case of two sets of
measurements and to the case of Gaussian and binary responses. We
developed and compared various inference schemes for multi-way
blockmodels, including Monte Carlo Markov chains and variational Bayes.

We further explored the impact of \textit{thresholding} and
\textit{binning} on the analysis. These censoring mechanisms are often
used as preprocessing steps. The transformed data are then amenable to
the analysis of coordination using off-the-shelf methods, including
Bayesian networks and popular blocking algorithms from the data mining
literature [\citet{bradbraurabitroy2009,ChakPapaModhFalo2004}].
The sensitivity analysis suggests that the impact of preprocessing
steps that involve censoring is substantial, both from a quantitative
perspective and in terms of its impact on biological discovery, in our
case study.

\section*{Acknowledgments}

The authors thank David Madigan for suggesting extensions of the mixed
membership blockmodel, in the
context of the analysis of adverse events.
EMA is an Alfred P. Sloan Research Fellow.
%

\begin{supplement}
\stitle{Supplement to ``Multi-way blockmodels for analyzing coordinated high-dimensional responses''}
\slink[doi]{10.1214/13-AOAS643SUPP}
\sdatatype{.pdf}
\sfilename{aoas643\_supp.pdf}
\sdescription{We provide additional supporting plots that show both
good and poor performance of the Hill estimator for the index of
regular variation in a variety of examples.}
\end{supplement}




\printaddresses
\end{document}